\documentclass[twocolumn,pra,showpacs,floatfix]{revtex4}
\usepackage{amsmath}
\usepackage{amsfonts}
\usepackage{amsthm}
\usepackage{graphicx}

\newcommand{\ket}[1]{\ensuremath{|#1\rangle}}
\newcommand{\degree}[1]{\ensuremath{#1^\circ}}
\newcommand{\CNOT}{\textsc{cnot}}
\newcommand{\NOT}{\textsc{not}}
\newcommand{\us}[1]{\ensuremath{#1\,\mu\text{s}}}

\begin{document}
\title{Compiling gate networks on an Ising quantum computer}
\author{M.~D.~Bowdrey}\email{mark.bowdrey@physics.org}
\affiliation{Centre for Quantum Computation, Clarendon Laboratory,
University of Oxford, Parks Road, OX1 3PU, United Kingdom}
\author{J.~A.~Jones}\email{jonathan.jones@qubit.org}
\affiliation{Centre for Quantum Computation, Clarendon Laboratory,
University of Oxford, Parks Road, OX1 3PU, United Kingdom}
\author{E.~Knill}\email{knill@boulder.nist.gov}
\affiliation{Mathematical and Computational Sciences Division,
National Institute of Standards and Technology, Boulder CO 80305}
\author{R.~Laflamme}\email{laflamme@iqc.ca}
\affiliation{Institute for Quantum Computing, University of
Waterloo, ON, N2L 3G1, Canada, and Perimeter Institute for
Theoretical Physics, 31 Caroline Street North, Waterloo, ON, N2L 2Y5,
Canada}
\date{\today}
\begin{abstract}
Here we describe a simple mechanical procedure for compiling a
quantum gate network into the natural gates (pulses and delays) for
an Ising quantum computer.  The aim is not necessarily to generate
the most efficient pulse sequence, but rather to develop an
efficient compilation algorithm that can be easily implemented in
large spin systems.  The key observation is that it is not always
necessary to refocus all the undesired couplings in a spin system.
Instead the coupling evolution can simply be tracked and then
corrected at some later time.  Although described within the
language of NMR the algorithm is applicable to any design of quantum
computer based on Ising couplings.
\end{abstract}
\maketitle

Quantum computers \cite{BennetDiv} have generated considerable
interest in recent years due to their apparent ability to perform
computations that are intractable on any classical computer.
Although the construction of a general purpose quantum computer
capable of solving real problems remains a challenge, preliminary
results have been demonstrated in several systems \cite{fortbook},
most notably nuclear magnetic resonance (NMR) \cite{cory96, cory97,
jones00, vandersypen01} and trapped ions \cite{monroe95, gulde03,
riebe04, barrett04}.  When implementing simple quantum algorithms on
devices with a small number of qubits, it is perfectly possible to
design an implementation ``by hand''.  With larger systems, however,
this approach becomes impractical, and it is desirable to automate
the process.  Here we describe a simple procedure that allows a
sequence of abstract quantum logic gates (the most common
device-independent description of a quantum algorithm) to be
compiled into a sequence of pulses and delays, which are the natural
gates on an Ising quantum computer.

By an Ising quantum computer we mean a system with a background
Hamiltonian
\begin{equation}
\mathcal{H}/\hbar= 2{\pi}\sum_i \nu_i I_z^i
+{\pi}\sum_{i<j}J_{ij}\,2{I_z^i}{I_z^j}
\end{equation}
where, following NMR notation \cite{ernst87}, $I_z=\sigma_z/2$ is
the angular momentum operator in the $z$ direction, $\nu_i$ is the
precession frequency of the $i$th qubit, and $J_{ij}$ is the Ising
coupling strength between pairs of qubits; here we assume that all
the $n(n-1)/2$ Ising couplings in an $n$ qubit system are of
significant size. This Hamiltonian arises naturally in NMR systems,
where the qubits correspond to spin-half particles and the scalar
$J$ coupling takes the Ising form within the weak coupling limit,
but also occurs more generally. It can be simplified by working in a
multiply rotating frame, where the frame used to describe each qubit
rotates at the Larmor precession frequency $\nu_i$, so that the
effective Hamiltonian contains only the coupling terms.  We also
assume that the system can be controlled by the application of
arbitrary single-qubit gates to any qubit or group of qubits. In an
NMR system these single-qubit gates would be implemented using
resonant radiofrequency pulses, with individual qubits being
distinguished by their unique Larmor frequencies.  We refer to these
single-qubit gates as \textit{pulses} whether or not they are
implemented in this way.

It is well known that any quantum algorithm can be implemented using
only single-qubit and two-qubit gates, such as \CNOT\
(controlled-\NOT) gates \cite{Barenco:1995a, DiVincenzo98}.
Furthermore, the \CNOT\ gate can itself be decomposed in terms of
(single-qubit) Hadamard gates and a controlled-$\sigma_z$ gate,
which converts $\ket{11}$ to $-\ket{11}$ while leaving other states
unchanged.  Finally, a controlled-$\sigma_z$ is essentially
equivalent to a \degree{90} evolution under an Ising coupling,
differing only by single-qubit $z$ rotations \cite{jones00}.  These
are most conveniently handled not by rotating the qubit, but rather
by rotating its reference frame, a method usually known as abstract
reference frames \cite{knill00}.  Similar techniques can be used to
implement more general controlled phase-shift gates , which are
useful for building such gates as the controlled
square-root-of-\NOT\ \cite{jones00}.  Thus arbitrary quantum
circuits can be constructed out of pulses and controlled evolutions
under the Ising Hamiltonian. Here we simply assume that a circuit
comprising only single-qubit gates and controlled phase-shift gates
is available.

In order to achieve the desired evolutions it is necessary to sculpt
the Hamiltonian into a more suitable form, in which only the desired
couplings of the required strength are present. This is easily
achieved by using spin echoes. Spin echoes are widely used in
conventional NMR experiments \cite{ernst87} to refocus undesirable
single-spin interactions, but they can also be used to effectively
rescale Ising coupling terms within a Hamiltonian.

\begin{figure}
\includegraphics{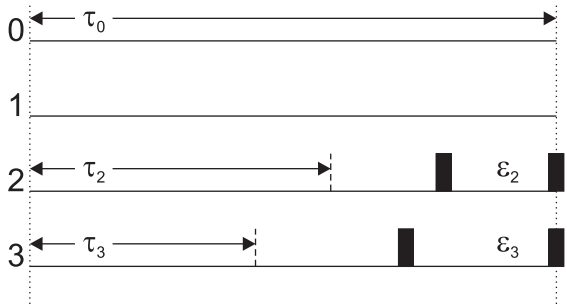}
\caption{An example of rescaling couplings between four qubits.
 The solid blocks indicate \NOT\ gates (\degree{180} pulses).}\label{fig:refocus}
\end{figure}

The basic procedure, which is based on the concept of an
\textit{average} Hamiltonian, is easily understood and is
exemplified by Fig.~\ref{fig:refocus}. The coupling between qubits
$0$ and $1$ evolves for the entire period, $\tau_0$, and so the
average coupling strength experienced by these qubits is equal to
the underlying coupling strength, $J_{01}$. The coupling from qubit
$0$ to qubit $2$ is partially refocused by applying a pair of \NOT\
gates (\degree{180} pulses) to one of the two qubits involved. The
effect of these \NOT\ gates is to reverse the coupling evolution
\cite{jones99, leung00b} for a period of length $\epsilon_2$, and so
it appears that the coupling has evolved only for a time
$\tau_2=\tau_0-2\epsilon_2$.  Alternatively (and equivalently) the
situation can be described as if the coupling had evolved for the
whole time $\tau_0$ but with a reduced coupling strength
$J'_{02}=(\tau_2/\tau_0)\,J_{02}$.  In the same way the coupling
between qubits $0$ and $3$ has been rescaled by a factor of
$\tau_3/\tau_0$, with $\tau_3=\tau_0-2\epsilon_3$. The effective
coupling strengths between other pairs of qubits can be worked out
in a similar way. Clearly $J'_{12}=(\tau_2/\tau_0)\,J_{12}$ and
$J'_{13}=(\tau_3/\tau_0)\,J_{13}$, by analogy with the couplings to
qubit $0$.  The last problem is to calculate the net evolution for
the coupling between qubits $2$ and $3$, where \NOT\ gates are
applied to \textit{both} qubits, and a little thought shows that the
scaling factor is $\tau_{23}/\tau_0$ with a net evolution time
$\tau_{23}=\tau_0-2(\epsilon_3-\epsilon_2)$.

The standard approach to date has been to use refocusing networks to
sculpt the underlying Ising Hamiltonian into some ideal form,
usually retaining only a small number of coupling interactions. In
particular, efficient refocusing schemes are known \cite{jones99,
leung00b} to retain any one coupling in an extended network. There
are, however, some limitations on the average Hamiltonians that can
be achieved. For example, consider a system of three coupled spins,
with a Hamiltonian containing three coupling terms: it is possible
to keep any one of these terms while refocusing the other two, but
it is not possible to keep two terms while refocusing the third.  If
such an average Hamiltonian is desired it is usually necessary to
implement the two Hamiltonians sequentially. Designing evolution
networks in large systems can become a complex business.

Our compilation algorithm provides an alternative approach for
designing pulse sequences on Ising quantum computers with large
numbers of spins.  The aim is not to generate the most efficient
pulse sequence, but rather to develop a simple algorithm that can be
easily implemented in large spin systems. The key observation is
that it is not, in fact, necessary to implement the exact Ising
coupling evolutions shown in the network.  In particular, it is not
always necessary to refocus all the undesired Ising couplings in a
spin system. Instead the coupling evolution can simply be tracked,
and their values corrected at some appropriate time. This is because
single-qubit gates applied to one qubit commute with couplings that
do not involve this qubit.  Thus it is only necessary to achieve the
``correct'' evolution for those qubits to which single-qubit gates
are applied. Furthermore, since single-qubit gates applied to
different qubits all commute with one another, two or more
``simultaneous'' single-qubit gates can in fact be applied
sequentially in any order.

The first stage is to redraw the network so that single-qubit gates
are applied sequentially, rather than in parallel. Simultaneous
gates can be applied in any order; the simplest approach is to apply
them from top to bottom. We refer to the qubit to which a
single-qubit gate is applied as the \textit{target} qubit, and any
other qubits as \textit{control} qubits.  It is \textit{not}
necessary that evolution periods generate any particular coupling
evolution between the control qubits, but these additional
evolutions must be tracked through the circuit.

Couplings are tracked by recording the net evolution angle (modulo
\degree{360}) generated for each of the $n(n-1)/2$ couplings at all
significant points in the network, where the change in the evolution
angle is given by
\begin{equation}
\delta\theta_{ij}=\pi J'_{ij} \tau,
\end{equation}
and $J'_{ij}$ is the average coupling strength between qubits $i$
and $j$ during a time period $\tau$.  It is essential that all the
coupling angles involving a target qubit have the correct value as
defined by the gate network before applying a single-qubit gate to
the target. This is easily achieved by applying \NOT\ gates to the
control spins, thereby changing the average coupling strengths, so
that all the couplings reach the desired angle in the same evolution
time. When the single-qubit gate is applied to the target qubit, all
coupling angles to this qubit are reset to zero. Couplings between
control qubits are simply tracked throughout the process.

\begin{figure}
\includegraphics{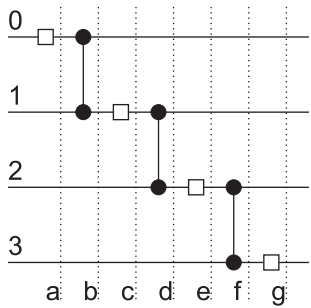}
\caption{An example quantum circuit, based on the algorithmic
benchmark of Knill \textit{et al.} \cite{knill00}.  White squares
indicate \degree{90_y} pulses (pseudo-Hadamard gates) while black
circles connected by control lines are Ising coupling gates with a
target angle of \degree{90} between the coupled
qubits.}\label{fig:circuit}
\end{figure}

\begin{figure}
\includegraphics{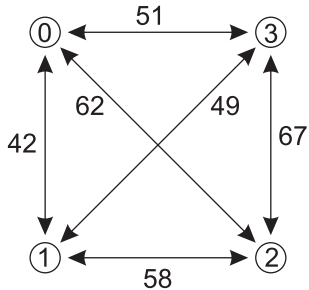}
\caption{An example Ising quantum computer, described by indicating
the strengths of the Ising couplings (measured in Hz) between all
pairs of qubits; these couplings were chosen
arbitrarily.}\label{fig:molecule}
\end{figure}

This procedure can be clarified by considering a concrete example,
and here we consider the implementation of the quantum circuit shown
in Fig.~\ref{fig:circuit} on the Ising quantum computer depicted in
Fig.~\ref{fig:molecule}.  The circuit is based on the algorithmic
benchmark of Knill \textit{et al.} \cite{knill00}, but the coupling
strengths shown in the Ising computer were chosen at random.  The
key values in an Ising computer are the coupling angles achieved
between pairs of qubits, and these will be considered at the points
marked with lower case letters in Fig.~\ref{fig:circuit}.

The first part of the circuit is simple: the system begins with all
coupling angles set to zero, and the first single-qubit gate has no
effect, so that at point (a) all coupling angles are \degree{0}.  By
point (b) it is necessary that the coupling angle between qubits 0
and 1 be increased to \degree{90}, while all other coupling angles
to qubit 1 (which is the target of the next single-qubit gate)
remain at zero; all further couplings can simply be tracked.  This
can be achieved by evolving under the coupling Hamiltonian for a
time $1/2J_{01}=\us{11905}$ with \NOT\ gates applied to qubits 2 and
3 at the middle and end of the coupling periods, acting to
completely suppress the couplings from these qubits to qubit 1.  The
couplings from the other control qubits (2 and 3) to qubit 0 are
also suppressed, \textit{but} the coupling between qubits 2 and 3
will evolve for the whole coupling time, and thus achieve a coupling
angle of \degree{144}. The next step, at point (c), is to apply a
single-qubit gate to qubit 1; at this point all coupling angles
involving this qubit are reset to zero, while leaving all other
angles unchanged. Thus the three coupling angles involving only
control qubits still have the values $\theta_{02}=\degree{0}$,
$\theta_{03}=\degree{0}$, and $\theta_{23}=\degree{144}$.  These
(and further) evolutions are summarized in Table~\ref{tab:angles}.
\begin{table}
\caption{Coupling evolution angles (measured in degrees, and rounded
to the nearest degree) between every pair of qubits at various
points in the initial part of the circuit shown in
Fig.~\ref{fig:circuit}. Coupling angles indicated in boldface are
coupling angles to target qubits that must be set correctly; other
angles are simply being tracked as part of the
algorithm.}\label{tab:angles}
\begin{center}
\begin{tabular}{l|rrrrrrrr}
angle&\makebox[2em][r]{a}&\makebox[2em][r]{b}&\makebox[2em][r]{c}&\makebox[2em][r]{d}&\makebox[2em][r]{e}&\makebox[2em][r]{f}&\makebox[2em][r]{g}\\\hline
0,1       &0&\textbf{90}&  0&70         &70&70         &70\\
0,2       &0& 0         &  0&\textbf{ 0}& 0&276        &276\\
0,3       &0& 0         &  0& 0         & 0&\textbf{ 0}& 0\\
1,2       &0&\textbf{ 0}&  0&\textbf{90}& 0&78         & 78\\
1,3       &0&\textbf{ 0}&  0&76         &76&\textbf{ 0}& 0\\
2,3       &0&144        &144&\textbf{ 0}& 0&\textbf{90}& 0\\
\end{tabular}
\end{center}
\end{table}

The first interesting point in the algorithm occurs at (d).  At this
stage the algorithm requires the coupling angles to qubit 2 to be
set correctly, that is $\theta_{02}=\degree{0}$,
$\theta_{12}=\degree{90}$, and $\theta_{23}=\degree{0}$.  Allowing
for the current values of the coupling angles, and recalling that
coupling angles are defined modulo \degree{360}, the additional
evolution required is $\delta_{02}=\degree{0}$,
$\delta_{12}=\degree{90}$, and $\delta_{23}=\degree{216}$.
Considering these evolutions individually, the coupling times
required are $\tau_{02}=\us{0}$, $\tau_{12}=\us{8621}$ and
$\tau_{23}=\us{17910}$.  Clearly the last coupling requires the
longest time, and so all the desired coupling angles can be achieved
by evolving under the full coupling Hamiltonian for \us{8955}, with
two couplings, $J_{02}$ and $J_{12}$, being (respectively) totally
and partially refocused.  Thus \NOT\ gates should be applied to
qubit 0 at the middle and end of the coupling period, while the
first \NOT\ gate is applied to qubit 1 after a time
$\tau_{12}+(\tau_{23}-\tau_{12})/2=\us{13266}$, with the second
\NOT\ gate applied at the end. The remaining coupling angles can
then be calculated by noting that $J_{03}$ is completely refocused,
$J_{13}$ evolves for the same time as $J_{12}$, giving an angle of
\degree{76}, and $J_{01}$ evolves for a net time of \us{9288},
giving an additional coupling angle of \degree{70}.

\begin{figure*}
\includegraphics{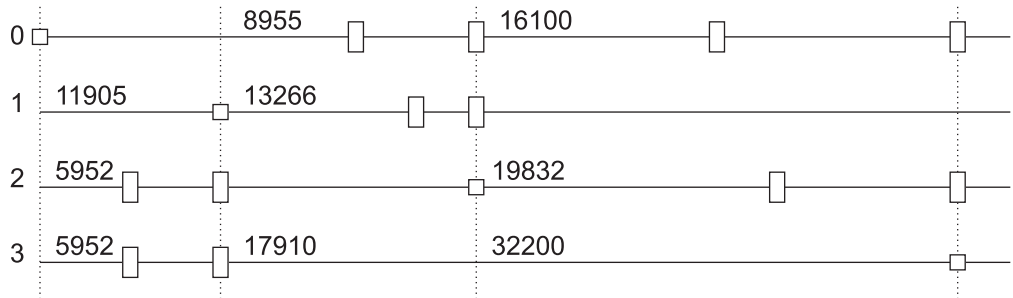}
\caption{Sequence of refocusing pulses used to implement the circuit
shown in Fig.~\ref{fig:circuit} on the Ising quantum computer shown
in Fig.~\ref{fig:molecule}.  White squares indicate \degree{90_y}
pulses as before, while white rectangles indicate \NOT\ gates. These
single-qubit gates are assumed to be instantaneous.  The evolution
periods are drawn approximately to scale, and the lengths of some
periods (measured in $\mu\text{s}$) are indicated.}
\label{fig:pulses}
\end{figure*}
At point (e), a single-qubit gate is applied to qubit 2, and so all
coupling angles involving this qubit are reset to zero.  The
couplings at point (f) are then set using the same method as was
used for (d). This time the limit coupling is $J_{13}$ which
requires a time of $\tau_{13}=\us{32200}$ to acquire an additional
coupling angle of $\delta_{13}=\degree{284}$.  The coupling $J_{03}$
is completely refocused by applying \NOT\ gates to qubit 0 at the
middle and end of this coupling period, while the coupling $J_{23}$
must be partially refocussed by applying the first \NOT\ gate to
qubit 2 after a time of \us{19832}.  The remaining couplings are
tracked as usual, and their values are shown in
Table~\ref{tab:angles}.  The complete sequence of refocusing pulses
and single-qubit gates is shown in Fig.~\ref{fig:pulses}. Note that
the evolution times depicted are those calculated above, which
include the effects of rounding errors; exact calculations would
give slightly different times.

As discussed above, this algorithm is not necessarily intended to
produce the most efficient pulse sequence, but it is still necessary
to check that the simplicity of implementation is not achieved at an
excessive cost either in pulse sequence length or in the number of
pulses required. Clearly the complexity of a pulse sequence will
depend on details of the circuit being implemented and the range of
coupling strengths in the experimental system. It is, however,
simple to consider some limiting situations.

Any portion of a circuit in a system of $n$ qubits can be classified
according to the number $p$ of target qubits involved, which
obviously lies between the extreme values of $1$ and $n$. Using our
new algorithm this section will be divided into at most $p$ separate
sections, each of which will require $2(n-2)$ refocusing pulses; in
total the implementation will require less than
$2\times{n}\times{p}$ pulses. Using the Hadamard matrix approach
will require between $1$ and $p$ refocusing periods (depending on
exactly which couplings have to be controlled), each of which will
contain approximately $n^2$ refocusing pulses \cite{leung00b}. The
relative efficiency of our new algorithm against the Hadamard
algorithm will depend on the number of refocusing periods required
by the latter.  In a system such as that shown in
Fig.~\ref{fig:molecule}, where all the coupling strengths are
different, it will not be possible in general to combine different
evolution periods, and so the Hadamard approach will require about
${p}\times{n^2}$ pulses. Thus our new algorithm will normally use
fewer pulses than previous methods. The time required to implement a
set of gates is another important consideration, but once again will
vary greatly from network to network.  It is easy to see that pulse
sequences produced by our new algorithm will in the worst case take
$p$ times longer than those produced by current methods, although it
seems likely that the relative performance will be better than this
in real situations.

It is possible to imagine a large number of ways in which this
algorithm could be extended, resulting in simpler or shorter pulse
sequences. Here we confine ourselves to two particularly simple
extensions, both of which can be easily implemented.  Firstly, we
note that it is not in fact necessary to place our pairs of \NOT\
gates at the end of the coupling period, as shown in
Fig.~\ref{fig:refocus}.  Instead the pair of refocusing gates can be
placed at any point within the coupling period without affecting the
net coupling to the target spin (the position will, of course,
affect the net couplings to other control spins).  In particular, we
can choose to place the first \NOT\ gate at the beginning of a
coupling period, and if this period is preceded by one with a \NOT\
gate at the end then two of the four \NOT\ gates can be canceled.

Secondly, as described so far, it has been necessary to set coupling
evolution angles to a target spin to the correct value modulo
\degree{360}; thus qubit pairs can, in principle, be required to
undergo evolutions of up to \degree{360} to achieve the desired
angle. This is, however, excessive.  We note that
\begin{equation}
\exp[-i\pi\,2I^j_zI^k_z]=\exp[-i\pi(I^j_z+I^k_z)]
\end{equation}
and so a \degree{180} coupling evolution on a pair of qubits is
equivalent to a \degree{180} frame rotation applied to
\textit{both} qubits.  Thus it is never necessary for qubits to
undergo an evolution through more than \degree{180}, as any
additional evolution can be achieved through frame rotations.

Thirdly, we note that, in addition to partly refocusing coupling
periods, \NOT\ gates can also be used to negate the sign of an Ising
coupling.  If a pair of \NOT\ gates is applied to a control qubit at
the beginning and end of an evolution period, then the evolution of
the Ising coupling will be reversed throughout this period.  If
\NOT\ gates are already applied to the qubit during this period to
partially refocus a coupling, then two of the four \NOT\ gates can
be cancelled, so there will be no overall increase in the number of
gates.  Combined with the previous observation, we note that any
desired evolution can be achieved using a coupling evolution in the
range \degree{\pm90}.  This allows the implementation time to be
reduced by a factor of around four, at the cost of a small increase
in the number of \NOT\ gates.

Finally, we note that it is possible to combine the traditional
approach and our new algorithm to produce hybrid schemes.  For
example, it might be useful to compile portions of a network
independently of any larger computation, in effect treating them as
subroutines. To do this, it is necessary to ensure that \textit{all}
the couplings in the system have correct evolution angles at the end
of the subroutine, not just those involving the final target qubit.
This can be achieved by ending the network with a conventional
refocusing period.  If, however, the whole network is compiled
together, then this final refocusing period will normally not be
required.  Computations usually end with measurements on some of the
qubits in the computational basis, and these eigenstates commute
with the Ising Hamiltonian.  The results obtained from measurements
of these qubits are, therefore, independent of their couplings
angles, and it is not necessary to refocus couplings to them.

\begin{acknowledgments}
MB and JAJ thank the UK EPSRC for financial support.  RL would like
to thanks MITACS and ARDA.  Contributions
to this work by NIST, an agency of the US government, are not
subject to copyright laws.
\end{acknowledgments}

\end{document}